\documentclass[12pt,leqno]{amsart}
\usepackage[dvips]{graphicx}
\usepackage{amsfonts}
\usepackage{amssymb}
\usepackage{amsmath}
\usepackage{amscd}
\usepackage{graphicx}

\def\DATE{Yom Shishi 5 Tichri 5773: vendredi 21 septembre 2012}
\newtheorem{theorem}{Theorem}
\newtheorem{definition}[theorem]{Definition}
\newtheorem{corollary}[theorem]{Corollary}

\newtheorem{lemma}[theorem]{Lemma}

\newtheorem{proposition}[theorem]{Proposition}

\newcommand\C{\mathbb{C}}
\newcommand\R{\mathbb{R}}
\newcommand\g{\mathfrak{g}}

\newcommand\K{\mathbb{K}}

\newcommand\p{\mathcal{P}}

\newcommand\pf{\noindent{\it Proof. }}
\newcommand\ds{\displaystyle}
\email{elisabeth.remm@uha.fr}

\title{Associative and Lie deformations of Poisson algebras}
\author{Elisabeth Remm}
\date{Lundi 15 Octobre 2012}
\address{Universit\'{e} de Haute Alsace, LMIA, 4 rue des Fr\`{e}res Lumi\`{e}%
re, 68093 Mulhouse}

\begin{document}

\maketitle

\begin{abstract}
Considering a Poisson algebra as a nonassociative algebra satisfying the Markl-Remm identity, we study deformations of Poisson algebras as deformations of this nonassociative algebra. This gives a natural interpretation of deformations which preserve the underlying associative structure and of deformations which preserve the underlying Lie algebra.
\end{abstract}

\noindent{\it Keywords :} Poisson algebras, Deformations, Operads, Cohomology.

\noindent{\it MS classification numbers}: 17B63,  17Dxx, 53Dxx.

\section{Introduction}
The Poisson bracket is a multiplication which naturally appears when studying deformations of associative commutative algebras. For instance the algebra 
$\mathcal{C}^{\infty} (\mathbb{R}^2)$ with its ordinary multiplication $\mu=\mu_0$ admits a formal deformation $\sum_0^{\infty}t^n\mu_n$ such that the skew-symmetric bracket
$\left\{ a,b \right\} =\mu_1(a,b)-\mu_1(b,a)$ is the classical Poisson bracket (recalled in Section \ref{Generalities}). This deformation is connected
to the $*$-product and then to the theory of deformation quantization (see Section 1 of \cite{kontsevich}). 
This naturally leads to study deformations of Poisson algebras. But a Poisson algebra is usually defined by two multiplication, an associative commutative one $a\bullet b$
and a Lie bracket $\left\{a , b \right\}$ (also called Poisson bracket) which are linked by the Leibniz rule 
$\left\{ a \bullet b , c \right\}= a \bullet\left\{ b , c \right\}+\left\{ a , c \right\}\bullet b .$ The deformations of Poisson algebras which are
classically considered consist of those deforming the Lie bracket while the associative product remains unchanged. The first studied Poisson algebras were defined on 
associative algebras of functions whose product is undeformable. This explains why this type of deformations, that we call Lie deformations of Poisson algebras, 
were first studied. They are parametrized by the Poisson-Lichnerowicz cohomology. Here we want to give a general approach of deformations of Poisson algebras , that is, we make deformations where both products are deformed. We then use the presentation of Poisson algebras in \cite{markl-eli-poisson} with a single nonassociative multiplication which capture all informations. Then we find the Lie deformations as a particular case of deformations of this single multiplication but also the associative deformations obtained by deforming the associative product and letting the Lie bracket unchanged. We call Poisson-Hochschild the cohomology parametrizing the associative deformations 
and we define it in Section ??????.
We then describe the Poisson cohomology parametrizing the general deformations of Poisson algebras and study the interactions between Poisson, Poisson-Lichnerovicz and 
Poisson-Hochschild colomologies.

\section{Generalities on Poisson algebras} \label{Generalities}

\subsection{Definition}
Let $\K$ be a  field of characteristic $0$. A $\K$-Poisson algebra is a $\K$-vector space  $\p$ equipped with two bilinear products denoted by $x\bullet y$  and $\{x ,y \}$, with the following properties:
\begin{enumerate}
\item The couple $(\p,\bullet)$ is an associative commutative $\K$-algebra.

\item The couple $(\p, \{ , \})$ is  a $\K$-Lie algebra.

\item The products $\bullet$ and $\{, \}$ satisfy the Leibniz rule:
$$\{x\bullet y,z\}=x \bullet\{y,z\}+\{x,z\}\bullet y$$
for any $x,y,z \in \p.$
\end{enumerate}
The product $\{,\}$ is usually called Poisson bracket and the Leibniz identity means that the Poisson bracket acts as a derivation 
of the associative product.

\medskip
\noindent Classical examples: Poisson structures on the polynomial algebra. The polynomial algebra $A_n = \C[x_1, \cdots, x_n]$ is provided with several  Poisson algebra structures. 
These examples are well studied, see, for example, \cite{Dufour, Nico, Vaisman} for results on classification, 
or \cite{Pich} for the study of the Poisson-Lichnerowicz cohomology.

\subsection{ Non standard example: Poisson algebras defined by a contact structure.}
The first Poisson structures appeared in classical mechanics. In 1809 Sim\'eon Denis Poisson
introduced a bracket in the algebra of smooth functions on $\R^{2r}$:
$$ \{f, g\} = \ds \sum_{i=1}^r\left(\frac{\partial f}{\partial p_i}\frac{\partial g}
{\partial q_i}-\frac{\partial f}{\partial q_i}\frac{\partial g}{\partial p_i}\right).$$
This classical example has a natural generalization in symplectic geometry (\cite{godbillon}): Let $(M^{2p},\theta)$ be a symplectic manifold. For any Pfaffian form $\alpha$ on $M^{2p}$, we will denote by $X_\alpha$ the vector field defined by $\alpha=i(X_\alpha)\theta$, where $i(X)$ is the interior product by $X$: $(i(X)\theta)(Y)=\theta(X,Y)$. The Poisson bracket of two Pfaffian forms $\alpha,\beta$ on $M^{2p}$ is the Pfaffian form $\{\alpha,\beta\}=i([X_\alpha,X_\beta])\theta.$ If $\mathcal{D}(M^{2p})$ denotes 
the associative commutative algebra of smooth functions on $M^{2p}$, we provide it with a Poisson algebra structure letting $\{f,g\}=-\theta(X_{df},X_{dg}).$ This Poisson bracket satisfies $d(\{f,g\})=\{df,dg\}$.

We can also define a Poisson bracket in contact geometry (\cite{GRcontact}).  Let $(M^{2p+1},\alpha)$ be a contact manifold,
 that is, $\alpha$ is a Pfaffian form on the $(2p+1)$-dimensional differential manifold $M^{2p+1}$ 
satisfying $(\alpha \wedge (d\alpha)^p)(x) \neq 0$ for any $x \in M^{2p+1}.$ 
There exists one and only one vector field $Z_\alpha$ on $M^{2p+1}$, called the Reeb vector field of 
$\alpha,$ such that $\alpha(Z_\alpha)=1$ and $i(Z_\alpha)d\alpha=0$ at any point of $M^{2p+1}$. 
Let $\mathcal{D}_\alpha(M^{2p+1})$ be the set of first integrals of $Z_\alpha,$ that is,
    $$\mathcal{D}_\alpha(M^{2p+1})=\{f \in \mathcal{D}(M^{2p+1}), \ Z_\alpha (f)=0\}.$$
    Since we have $Z_\alpha(f)=i(Z_\alpha)df=0$, then $df$ is invariant by $Z_\alpha$.
    \begin{lemma}
    $\mathcal{D}_\alpha(M^{2p+1})$ is a  commutative associative subalgebra of $\mathcal{D}(M^{2p+1})$.
\end{lemma}
\pf This is a consequence of the classical formulae
$$Z_\alpha(f+g)=Z_\alpha(f)+Z_\alpha(g) \ {\rm and}  \ Z_\alpha(fg)=(Z_\alpha(f))g+f(Z_\alpha(g)).$$
\begin{lemma}
For any non zero Pfaffian form $\beta$ on $M^{2p+1}$ satisfying $\beta(Z_\alpha)=0$, 
there exists a vector field $X_\beta$ with $\beta(Y)=d\alpha (X_\beta,Y)$ for any vector field $Y$. 
Two vector fields $X_\beta$ and $X'_\beta$ with this property satisfy $i(X_\beta-X'_\beta)d\alpha=0$.
\end{lemma}
This means that $X_\beta$ is uniquely defined up to a vector field belonging to the distribution 
given by the characteristic space of $d\alpha$,
$$A(d\alpha)_x=\{X_x \in T_xM^{2p+1}, i(X_x) d\alpha(x)=0\}.$$ 
In any Darboux open set, the contact form writes as $\alpha=x_1dx_2+\cdots+x_{2p-1}dx_{2p}+dx_{2p+1}$. 
The Reeb vector field is $Z_\alpha=\partial/\partial x_{2p+1}$ and the form $\beta$ satisfying $\beta(Z_\alpha)=0$
 writes as $\beta=\sum_{i=1}^{2p}\beta_idx_i.$ Then we have
$$X_\beta=\sum_{i=1}^{p}(\beta_{2i}\partial/\partial x_{2i-1}-\beta_{2i-1}\partial/\partial x_{2i}).$$
For any $f\in \mathcal{D}_\alpha(M^{2p+1})$, we writes $X_f$ for $X_{df}$.
\begin{theorem}
The algebra  $\mathcal{D}_\alpha(M^{2p+1})$ is a Poisson algebra.
\end{theorem}
\pf (see \cite{GRcontact}). Let $f_1,f_2$ be in $\mathcal{D}_\alpha(M^{2p+1})$. Since we have $$d\alpha(X_{f_1},X_{f_2})=d\alpha(X_{f_1}+U_1,X_{f_2}+U_2)$$ for any $U_1,U_2 \in A(d\alpha)$, the bracket
$$\{f_1,f_2\}=d\alpha(X_{f_1},X_{f_2})$$
is well defined. It is a Poisson bracket.

\subsection{Poisson algebra viewed as nonassociative algebra}

In \cite{markl-eli-poisson}, we prove that any Poisson structure on a $\K$-vector space is also given by a nonassociative product denoted by $xy$ and satisfying the nonassociative identity
\begin{eqnarray}
\label{associator} 3A(x,y,z)=(x z) y+(y z) x-(y x) z-(z x) y,
\end{eqnarray}
where $A(x,y,z)$ is the associator $A(x,y,z)=(xy)z-x(yz)$. In fact, if $\p$ is a Poisson algebra with associative product $x\bullet y$ and  Poisson bracket $\{x,y\}$, then $xy$ is given by $xy=\{x,y\}+x \bullet y$. Conversely, the Poisson bracket and the associative product of $\p$ are the
 skew-symmetric part and the symmetric part of the product $xy$. Thus it is equivalent to present a Poisson algebra classically or by this nonassociative product.

If $\p$ is a Poisson algebra given by the nonassociative product (\ref{associator}), we denote by $\g_{\p}$ the Lie algebra on the same vector space $\p$ whose Lie bracket is $\{x,y\}=\frac{xy-yx}{2}$ and by $\mathcal{A}_{\p}$ the commutative associative algebra, on the same vector space, whose product is $x \bullet y=\frac{xy+yx}{2}$.

In \cite{Mic-Eli-Poisson}, we have studied algebraic properties of the nonassociative algebra $\p$. In particular we have proved that this algebra is flexible, power-associative, and admits a Pierce decomposition.

\noindent{\bf Remark.} A class of Poisson algebras is already defined with a single noncommutative multiplication but
starting with a Jordan algebra. In \cite{Skosyrskii}, a noncommutative Jordan algebra is viewed as a Jordan commutative algebra $J$ with an additional
skew-symmetric operator $ \lbrack\, , \rbrack: J \times J \rightarrow J$ such that
$$\lbrack x^2,y\rbrack =2\lbrack x,y\rbrack\cdot x.$$
This definition is equivalent to consider only one multiplication satisfying
$$(xy)x-x(yx)=(x^2y)x-x^2(yx).$$
A particular class of such algebras for which $A^{(+)}$ is associative corresponds to Poisson algebras.

\subsection{ Classification of complex Poisson algebras of dimension $2$ and $3$}

 If $e$ is an idempotent of the associative algebra, then the Leibniz rule implies that it
is in the center of the Lie algebra corresponding to the Poisson bracket. 
In fact if $e$ satisfies $e \bullet e =e$, thus $\{e\bullet e,x\}=2e\bullet \{e,x\}=\{e,x\}$. 
But if $y$ is a non zero vector with $e \bullet y= \lambda y$, then
$$ (e \bullet e) \bullet y = e \bullet y = \lambda y = e \bullet (e \bullet y)= \lambda^2 y.$$
This gives $\lambda^2= \lambda$, that is, $\lambda =0$ or $1$. 
Since we have $e \bullet \{e,x\}=2^{-1}\{e,x\}$, the vector $\{e,x\}$ is zero for any $x$ and $e$ is in the center 
of the Lie algebra corresponding to the Poisson bracket. 
This remark  simplifies the determination of all possible Poisson brackets when the associative product is fixed.
In the following, we give the associative and Lie products  in a fixed basis $\{e_i\}$ and the null products or the products which are deduced by commutativity or 
skew-symmetry are often not written.

\noindent{\it Dimension 2}
$$
\begin{array}{ll|l|l}
& {\rm algebra} & {\rm associative \ product} &  {\rm Lie \  product} \\
 \bullet & \p_1^2 & e_1\bullet
e_i=e_i , i=1,2; \ \  e_2\bullet e_2=e_2 & \{e_i,e_j\}=0. \\
 & &  \\
\bullet & \p_2^2  & e_1\bullet e_i=e_i , i=1,2 & \{e_i,e_j\}=0. \\
 & & \\
\bullet & \p_3^2  & e_1\bullet e_1=e_2  & \{e_i,e_j\}=0. \\
& &  \\
\bullet & \p_4^2  & e_1\bullet e_1=e_1  & \{e_i,e_j\}=0.\\
& &  \\
\bullet & \p_5^2  & e_i\bullet e_j=0  & \{e_1,e_2\}=e_2.\\
& &  \\
\bullet & \p_6^2  & e_i\bullet e_j=0  & \{e_i,e_j\}=0.\\
\end{array}
$$
\medskip
\noindent{\it Dimension 3}
$$
\begin{array}{ll|l|l}
\medskip
 \bullet & \p_1^3 & e_1\bullet
e_i=e_i , i=1,2,3; \ \  e_2\bullet e_2=e_2; \ \ e_3\bullet e_3=e_3, & \{e_i,e_j\}=0. \\
\medskip
 \bullet & \p_2^3 & e_1\bullet
e_i=e_i , i=1,2,3; \ \  e_2\bullet e_2=e_2; \ \ e_3\bullet e_3=e_2-e_1, & \{e_i,e_j\}=0. \\
\medskip
 \bullet & \p_3^3 & e_1\bullet
e_i=e_i , i=1,2,3; \ \  e_2\bullet e_2=e_2  & \{e_i,e_j\}=0. \\
\medskip
 \bullet & \p_4^3 & e_1\bullet
e_i=e_i , i=1,2,3; \ \  e_3\bullet e_3=e_2 & \{e_i,e_j\}=0. \\
\medskip
 \bullet & \p_5^3 & e_1\bullet
e_i=e_i , i=1,2,3  & \{e_2,e_3\}=e_3.\\
\medskip
 \bullet & \p_6^3 & e_1\bullet
e_i=e_i , i=1,2,3  & \{e_i,e_j\}= 0.\\
\medskip
 \bullet & \p_7^3 & e_1\bullet
e_i=e_i , i=1,2;\ \  e_2\bullet e_2=e_2  & \{e_i,e_j\}=0.\\

\medskip
 \bullet & \p_8^3 & e_1\bullet
e_1=e_1  & \{e_2,e_3\}=e_3.\\
\medskip
 \bullet & \p_9^3 & e_1\bullet
e_1=e_1  & \{e_i,e_j\}= 0.\\
\medskip
 \bullet & \p_{10}^3& e_1\bullet
e_i=e_i , i=1,2\ \   & \{e_i,e_j\}=0.\\
\medskip
 \bullet & \p_{11}^3 & e_1\bullet
e_1=e_1, \ \ e_2\bullet
e_2=e_3  & \{e_i,e_j\}=0.\\
\medskip
\bullet & \p_{12}^3(b) & e_1\bullet
e_1=e_2  & \{e_1,e_3\}=e_2+be_3. \\
\medskip
\bullet & \p_{13}^3 & e_1\bullet
e_1=e_2  & \{e_1,e_3\}=e_3. \\
\medskip
\bullet & \p_{14}^3 & e_1\bullet
e_1=e_2  & \{e_i,e_j\}=0. \\
\medskip
\bullet & \p_{15}^3 & e_1\bullet e_1=e_2; \ \ e_1\bullet e_2=e_3  & \{e_i,e_j\}=0.\\
\medskip
\bullet & \p_{16}^3(a) & e_i\bullet
e_j=0  & \text{\rm any Lie algebra. }
\end{array}
$$

It is also possible to establish this classification in small dimension starting from the nonassociative product. We can use, for example, technics used in 
\cite{Mic-Eli-2algebres} where we classify all the complex $2$-dimensional algebras (and in particular the Poisson algebras).

\section{Deformations of Poisson algebras }

In this section we recall briefly the classical notion of formal deformations of a $\mathbb{K}$-algebra. These deformations are parametrized by a cohomology, called deformation cohomology, which is often difficult to define globally and to compute explicitly. But using the operadic approach, we can sometimes obtain this cohomology using the associated operad: when the operad is Koszul, which is the case for the operad associated to Poisson algebras. When the operad is non Koszul the operadic and deformation cohomologies differ and the last one is even more complicated to describe see \cite{marklRemmdef}. Using the Markl-Remm definition of a Poisson algebra, we describe the formal deformations. So in this section,  we mean by Poisson algebra a $\K$-algebra defined by a nonassociative product satisfying Identity (\ref{associator}).  

\subsection{Formal deformations of a Poisson algebra}

 Let $R$ be a complete local augmented ring
such that the augmentation $\varepsilon$ takes values in $\K$. If $B$ is an $R$-Poisson algebra, we consider the $\K$-Poisson algebra $\overline{B}=\K \otimes_R B$ given by $\alpha(\beta\otimes b)=\alpha\beta\otimes b$, with $\alpha,\beta \in \K$ and $b \in B$. It is clear that $\overline{B}$ satisfies (\ref{associator}). An $R$-deformation of a $\K$-Poisson algebra $A$ is an $R$-Poisson algebra $B$  with a $\K$-algebra homomorphism
$$\varrho: \overline{B}\rightarrow A.$$
A formal deformation of $A$ is an $R$-deformation with $R=\K[[t]]$, the local ring of formal series on $\K$. We assume also that $B$ is
an $R$-free module isomorphic to $R \otimes A$.

Let $\K[\Sigma_3]$ be the $\K$-group algebra of the symmetric group $\Sigma_3$. We denote by $\tau_{ij}$ the transposition exchanging $i$ and $j$ and by $c$ the cycle $(1,2,3)$. Every $\sigma \in \Sigma_3$ defines a natural action on any $\K$-vector space $W$ by:
$$
\begin{array}{cccc}
\Phi_\sigma : &W^{\otimes ^3} & \longrightarrow  & W^{\otimes ^3}\\
& x_1\otimes x_2\otimes x_3 & \longrightarrow  & x_{\sigma(1)}\otimes x_{\sigma(2)}\otimes x_{\sigma(3)}.
\end{array}$$
We extend this action of $\Sigma_3$ to an action of the algebra $\K[\Sigma_3]$. If $v=\Sigma_i a_i\sigma _i \in \K[\Sigma_3]$, then
$$\Phi_v=\Sigma_i a_i\Phi_{\sigma _i}.$$
Consider $v_{P}$ the vector of $\K[\Sigma_3]$
$$v_{P}=3Id-\tau_{23}+\tau_{12}-c+c^2.$$
Let $\p$ be a Poisson algebra and $\mu_0$ its (nonassociative) multiplication. Identity (\ref{associator}) writes as
$$(\mu_0\circ_1 \mu_0)\circ \Phi_{v_P}-3(\mu_0\circ_2 \mu_0)=0$$
where $\circ_1$ and $\circ_2$ are the $comp_i$ operations given by
$$
\begin{array}{l}
(\mu\circ_1 \mu')(x,y,z)=\mu(\mu'(x,y),z),\\
(\mu\circ_2 \mu')(x,y,z)=\mu(x,\mu'(y,z))\\
\end{array}
$$
for any bilinear maps $\mu$ and $\mu'$.
\begin{theorem}
A formal deformation $B$ of the $\K$-Poisson algebra $A$ is given by a family of linear maps
$$\{\mu_i: A \otimes A \rightarrow A, \ i \in \mathbb{N}\}$$
satisfying
\begin{enumerate}
\item $\mu_0$ is the multiplication of $A$,
\item $(D_k)$: $\sum_{i+j=k,\ i,j\geq 0} \ (\mu_i\circ_1 \mu_j)\circ \Phi_{v_P}=3\sum_{i+j=k,i,j\geq 0} \ \mu_i\circ_2 \mu_j$ for each $k \geq1$.
\end{enumerate}
\end{theorem}
\pf  The multiplication in $B$ is determined by its restriction to $A \otimes A$
( \cite{doubekMarkl}). We expand $\mu(x,y)$ for $x,y$ in $A$ into the power series
$$ \mu(x,y)=\mu_0(x,y)+t\mu_1(x,y)+t^2\mu_2(x,y)+\cdots + t^n\mu_n(x,y)+\cdots$$
then $\mu$ is a Poisson product if and only if the family $\{\mu_i\}$ satisfies condition $(D_k)$ for each $k$.

\medskip

 \noindent{\bf Remark. } As $R$ is a complete ring, this formal expansion is convergent. 
It is also the case if $R$ is a valued local ring (see \cite{Mic-Eli-valued}).

\medskip
Let $\K=\C$ or an algebraically closed field. If $\{e_1,\cdots,e_n\}$ is a fixed basis of $\K^n$, we denote by $\p_n$ the set of all Poisson algebra structures on $\K^n$, that is, the set of  structure constants $\{\Gamma_{ij}^k\}$ given by $\mu(e_i,e_j)=\sum_{k=1}^n\Gamma_{ij}^ke_k.$ Relation (\ref{associator}) is equivalent to
$$
\displaystyle\sum_{l=1}^n 3\Gamma_{ij}^l\Gamma_{lk}^s-3\Gamma_{il}^s\Gamma_{jk}^l
-\Gamma_{ik}^l\Gamma_{lj}^s  -\Gamma_{jk}^l\Gamma_{li}^s
+\Gamma_{ji}^l\Gamma_{lk}^s+\Gamma_{ki}^l\Gamma_{lj}^s=0.
$$
Thus $\p_n$ is an affine algebraic variety. If we replace $\p_n$ by a differential graded scheme, 
we call {\it Deformation Cohomology,} the cohomology of the tangent space of this scheme.

\noindent{\bf Remark.} This cohomology of deformation is defined in same manner for any $\K$-algebra
and more generally for any $n$-ary algebra. If we denote by $H_{def}(A)=\oplus _{n\geq 0}H_{def}^n (A)$
the deformation cohomology of the algebra $A$, then $H_{def}^0(A)=\K$, $H_{def}^1(A)$ is the
space of outer derivations of $A$ and the coboundary operator $\delta_{def}^1$ corresponds to the operator of
derivation, and the space of $2$-cocycles is determined by the linearization of the identities defining $A$.
Thus, in any case, the three first spaces of cohomology are easy to compute. But  the determination of the
spaces $H_{def}^n(A)$ for $n \geq 3$ is usually not easy; we cannot deduce for example $H_{def}^3(A)$ directly from the knowledge of $H_{def}^2(A)$. 
However we have the following result:
\begin{proposition}
Let $\mathcal{P}_A$ be the quadratic operad related to $A$. If $\mathcal{P}_A$ is a Koszul operad, then
$H_{def}(A)$ coincides with the natural operadic cohomology.
\end{proposition}
For example, if $A$ is a Lie algebra or an associative algebra,
the corresponding operads $\mathcal{L}ie$ and $\mathcal{A}ss$ are Koszul and $H_{def}(A)$ 
coincides with the operadic cohomology, that is, respectively,
the Chevalley-Eilenberg cohomology and the Hochschild cohomology.
Examples of determination of $H_{def}(A)$ in the non-Koszul cases can be found in (\cite{Remmtartu, nicoeliJOA}).
A theory of deformations on non-Koszul 
operads in presented in (\cite{marklRemmdef}).

\subsection{The operadic cohomology of a Poisson algebra}

 Let $\mathcal{P}oiss$ be the quadratic binary operad associated with Poisson algebras. Recall briefly its definition.
Let
$E=\Bbb{K} \left[ \Sigma_2 \right]$ be the $\Bbb{K}$-group algebra
of the symmetric group on two elements. The basis of the free
$\mathbb{K}$-module $\mathcal{F}(E)(n)$ consists of the
"$n$-parenthesized products" of $n$ variables $\left\{ x_1, ...,x_n
\right\}$. Let $R$ be the $\Bbb{K}\left[
\Sigma_{3}\right]$-submodule of $\mathcal{F(}E)(3)$ generated by
the vector
$$u=3x_1(x_2x_3)-3(x_1x_2)x_3+(x_1x_3)x_2+(x_2x_3)x_1-(x_2x_1)x_3-(x_3x_1)x_2.$$
Then $\mathcal{P}oiss$ is the binary quadratic operad with generators $E$ and relations $R$.
It is given by
$$\mathcal{P}oiss(n)=
(\mathcal{F(}E)/\mathcal{R)}(n)=\frac{\mathcal{F(}E)(n)}{\mathcal{R(}n%
\mathcal{)}}
$$
where $\mathcal{R}$ is the operadic ideal of $\mathcal{F}(E)$ generated by
$R$ satisfying $\mathcal{R}(1)=\mathcal{R}(2)=0 $,
$\mathcal{R}(3)=R.$ The dual operad $\mathcal{P}oiss^!$ is equal
to $\mathcal{P}oiss$, that is, $\mathcal{P}oiss$ is self-dual. In
 \cite{G.R3} we defined, for a binary
quadratic operad $\mathcal{E}$, an associated quadratic operad
$\widetilde{\mathcal{E}}$ which gives a functor
$$\mathcal{E} \otimes \widetilde{\mathcal{E}}\rightarrow \mathcal{E}.$$ In the case $\mathcal{E}=\mathcal{P}oiss,$ we have
$\widetilde{\mathcal{E}}=\mathcal{P}oiss^!=\mathcal{P}oiss.$ All these properties show that the operad
$\mathcal{P}oiss$ is a Koszul operad (see also \cite{loday}). In this case the cohomology of deformation of $\mathcal{P}oiss$-algebras coincides with the natural operadic cohomology. An explicit presentation of the space of $k$-cochains
is given in \cite{M.S.S} :
$$\mathcal{C}^k(\mathcal{P},\mathcal{P})=
\mathcal{L}in(\mathcal{P}oiss(n)^{!}\otimes _{\Sigma _n}V^{\otimes n},V)=End(\mathcal{P}^{\otimes k},\mathcal{P})$$
where $V$ is the underlying vector space (here $\C^n$).
The cohomology associated with the complex $(\mathcal{C}^k(\mathcal{P},\mathcal{P}),\delta^k_P)_k$ where 
$\delta^k_P$ denotes the coboundary operator
$$\delta^k_P:\mathcal{C}^k(\mathcal{P},\mathcal{P})\rightarrow \mathcal{C}^{k+1}(\mathcal{P},\mathcal{P}),$$ is denoted by 
$H^*_P(\p,\p)$. We will describe the coboundary operators $\delta^2_P$ in Subsection \ref{4.2} and $\delta^k_P$ in Section
\ref{7}.

\noindent{\bf Consequence: The deformation cohomology of a Poisson algebra.} \label{4.2}
If $ \mathcal{P} $ is a Poisson algebra, then $H_{def}(\mathcal{P})$ is the operadic cohomology $H^*_P(\p,\p)$ or briefly 
$H^*_P(\p).$ 

\subsection{Some relations} 
 Let $\p$ be a Poisson algebra whose nonassociative product $\mu_0(X,Y)$ is denoted by $X \cdot Y$. Let
$\frak{g}_\mathcal{P}$ and $\mathcal{A}_{\mathcal{P}}$ be its
corresponding Lie and associative algebras. We denote by
$H_C^{*}(\frak{g}_\mathcal{P},\frak{g}_\mathcal{P})$ the Chevalley-Eilenberg
cohomology of $\frak{g}_{\mathcal{P}}$ and by
$H_H^{*}(\mathcal{A}_\mathcal{P},\mathcal{A}_\mathcal{P})$ the
Hochschild cohomology of $\mathcal{A}_{\mathcal{P}}$.
A important part of this work
devoted to describe the coboundary operator and its links with the Chevalley-Eilenberg and Hochschild cobounary operators.
We focus in this section on the degree 2 because it is related to the parametrization of deformations.
The condition $(D_1)$
 writes as
$$(\mu_0\circ_1 \mu_1+\mu_1\circ_1 \mu_0)\circ \Phi_{v_{P}}=3(\mu_0\circ_2 \mu_1+\mu_1\circ_2 \mu_0),$$
that is,
$$
\begin{array}{l}
3\mu_1(\mu_0( x,y),z)-3\mu_1 (x,\mu_0(y,z))-\mu_1( \mu_0( x,z),y)- \mu_1( \mu_0( y,z),x)\\
 +\mu_1( \mu_0( y,x),z)+\mu_1( \mu_0( z,x),y) +3\mu_0(\mu_1(x,y), z)-3\mu_0( x , \mu_1 (y,z))\\
 -\mu_0(\mu_1 (x,z) , y)- \mu_0(\mu_1 (y,z),x) + \mu_0(\mu_1 (y,x), z)
+\mu_0(\mu_1 (z,x), y)=0
\end{array}
$$
for any $x,y,z \in \p$.
If $\varphi$ is a $2$-cocycle of $H^2_{def}(\p)$, this implies
$$ \delta_P^2\varphi= (\varphi \circ_1\mu +\mu \circ_1 \varphi )\circ \Phi_{v_P}
-3(\varphi \circ_2\mu +\mu \circ_2 \varphi )\circ \Phi_{Id}.$$
Recall that $v_P=3Id-\tau_{23}+\tau_{12}-c+c^2.$
\medskip

\noindent Let $\varphi:\p^{\otimes^2}\rightarrow \p$ be a bilinear map
and $\mu$ be the nonassociative multiplication of the Poisson algebra $\p$. We denote by $\varphi _a=
\frac{\varphi -\widetilde{\varphi }}{2} $ and $\varphi _s=\frac{\varphi +\widetilde{\varphi }}{2} $
the skew-symmetric and symmetric parts of $\varphi $ with $\widetilde{\varphi }(X,Y)=\varphi (Y,X).$
We consider the following trilinear maps:
$$
\begin{array}{l}
\mathcal{L}_C (\varphi)
 =\frac{1}{2} [ \varphi \circ_1 \mu\circ \Phi _{Id+c+c^2-\tau _{12}-\tau _{13} -\tau _{23}}+
(\mu\circ_1\varphi -\mu \circ_2\varphi  )\circ \Phi _{Id+c+c^2}  ], \\
\\
\mathcal{L}_H(\varphi) =\frac{1}{2} [ \mu\circ _1\varphi \circ \Phi _{Id-c}  -
\mu\circ _2\varphi \circ \Phi _{Id-c^2}  +\varphi \circ_1 \mu\circ \Phi _{Id+\tau _{12}}-
\varphi \circ_2 \mu\circ \Phi _{Id+\tau _{13}} ].
\end{array}
$$
If $\varphi =\varphi _a$, that is, if $\varphi$ is skew-symmetric, then
$\mathcal{L}_C (\varphi_a)=\delta ^2_{C,\{ \, , \}}\varphi_a $ where $\delta ^2_{C,\{ \, , \}}$ is the
Chevalley-Eilenberg coboundary operator of the cohomology of the Lie algebra $\frak{g}_\p$ associated with $\p.$
Similarly if $\varphi =\varphi _s$, that is, if $\varphi$ is symmetric, then
$\mathcal{L}_H (\varphi_s)=\delta ^2_{H,\bullet}\varphi_s $ where $\delta ^2_{H,\bullet}$ is the
Hochschild coboundary operator of the cohomology of the associative algebra $\mathcal{A}_\p$ associated with $\p.$
Since no confusions are possible  we will  write $\delta _C^*$ and $\delta _H^*$ in place of $\delta^*_{C,\{ \, , \}}$
$\delta ^*_{H,\bullet}.$ Then for any bilinear map $\varphi $ on  $\p^{\otimes^2}$ with skew-symmetric part $\varphi _a$ and
symmetric part $\varphi _s,$ we obtain
\begin{itemize}
\item $4\delta^2 _C\varphi _a=(\mu \circ _1\varphi +\varphi \circ _1\mu
-\mu \circ _2\varphi +\varphi \circ _2\mu )\circ \Phi _V$ with $V=Id-\tau _{12}-\tau _{13} -\tau _{23}+c+c^2.$

\item $4\mathcal{L}_C(\varphi _s)=(\mu \circ _1\varphi -\mu \circ _2\varphi)\circ \Phi _W+( \varphi \circ _1\mu
+\varphi \circ _2\mu )\circ \Phi _V$
with $W=Id+\tau _{12}+\tau _{13} +\tau _{23}+c+c^2.$

\item $4\mathcal{L}_H(\varphi _a)=\mu \circ _1\varphi \circ \Phi _{Id-\tau _{12}+\tau _{13} -c}
+\mu \circ _2\varphi\circ \Phi _{-Id-\tau _{13}+\tau _{23}  +c^2}
+ \varphi \circ _1\mu \circ  \Phi _{Id+\tau _{12}+\tau _{13} +c}
+\varphi \circ _2\mu \circ \Phi _{-Id-\tau _{13}-\tau _{23}  -c^2}.$

\item $4\delta^2 _H\varphi _s=\mu \circ _1\varphi \circ \Phi _{Id+\tau _{12}-\tau _{13} -c}
+\mu \circ _2\varphi\circ \Phi _{-Id+\tau _{13}-\tau _{23}  +c^2}
+ \varphi \circ _1\mu \circ  \Phi _{Id+\tau _{12}-\tau _{13} -c}
+\varphi \circ _2\mu \circ \Phi _{-Id+\tau _{13}-\tau _{23}  +c^2}.$

\end{itemize}

At least we introduce the following operators, $\mathcal{L}_1$ which acts on the space of skew-symmetric bilinear maps
and $\mathcal{L}_2$ which acts on the space of symmetric bilinear maps on $\p:$

\begin{itemize}

\item $\begin{array}{lll}
4\mathcal{L}_1(\varphi_a)& =& \mu \circ_1 \varphi \circ \Phi_{ \tau_{13}-\tau_{23}-c+c^2  }
+\mu \circ_2 \varphi \circ\Phi_{-Id-\tau_{12}+\tau_{23}+c}+ \varphi\circ_1 \mu \circ\Phi_{Id+\tau_{12}} \\
&&+\varphi\circ_2 \mu \circ\Phi_{-\tau_{13}-c^2}.
\end{array}$

        \medskip

    \item $\begin{array}{lll}
4\mathcal{L}_2(\varphi_s)&=&\mu \circ_1 \varphi \circ \Phi_{ 2Id+2\tau_{12}-\tau_{13}-\tau_{23}-c-c^2  }
+\mu \circ_2 \varphi \circ\Phi_{-Id+\tau_{12}-\tau_{13}+c}\\
& & + \varphi\circ_1 \mu \circ\Phi_{Id+\tau_{12}+2\tau_{13}-4c}+\varphi\circ_2 \mu \circ\Phi_{-4Id+\tau_{13}+2\tau_{23}+c^2}.
\end{array}$
\end{itemize}

\begin{lemma}
$\mathcal{L}_1(\varphi_a)=0$ if and only if $\varphi _a$ is a skew derivation of the associative product associated with
$\mu,$ that is:
$$\varphi _a(x \bullet y,z)=x \bullet \varphi _a (y,z)+\varphi _a(x,z) \bullet y.$$
\end{lemma}

\pf
$$\begin{array}{l}
\medskip
\varphi _a(x \bullet y,z)-x \bullet \varphi _a (y,z)-\varphi _a(x,z) \bullet y \\
\medskip
\ \ \ \ = \frac{1}{2}(\varphi _a(x y+yx,z)-x \varphi _a (y,z)-\varphi _a (y,z)x-\varphi _a(x,z)  y -y\varphi _a(x,z))\\
\medskip
\ \ \ \ = \frac{1}{2} ( \varphi _a(x y,z)+\varphi _a(yx,z)+ x \varphi _a (z,y)+\varphi _a (z,y)x
+\varphi _a(z,x)  y +y\varphi _a(z,x))\\
\ \ \ \  = \frac{1}{4}\mathcal{L}_1(\varphi_a)(x,y,z).
\end{array}$$

\begin{proposition}  For every bilinear map $\varphi$ on $\p$, we have
\begin{eqnarray}
\label{deltaQ}\delta^2_P \varphi =2(\delta^2_C \varphi_a+
\mathcal{L}_C (\varphi_s)+\delta^2_H \varphi_s+\mathcal{L}_H (\varphi_a)+
\mathcal{L}_1(\varphi_a)+\mathcal{L}_2({\varphi_s})).
\end{eqnarray}
\end{proposition}
\medskip

\begin{corollary}
Let $\varphi $ be a bilinear map and $\varphi _a$ and $\varphi _s$
the skew-symmetric and the symmetric parts of $\varphi $. We have :

\begin{eqnarray}\label{chev}
12 \delta ^2_C\varphi _a =  \delta ^2_P \varphi \circ \Phi _{Id-\tau _{12}-\tau _{13} -\tau _{23}+c+c^2}
\end{eqnarray}
and
\begin{eqnarray}\label{chev1}
12 \delta ^2_H\varphi _s =  \delta ^2_P \varphi \circ \Phi _{Id-\tau _{13}+\tau _{23}-c^2}.
\end{eqnarray}
\end{corollary}

\section{Particular deformations: Lie and associative deformations of a Poisson algebra}

\subsection{Lie deformations}
\begin{definition}
We say that the formal deformation $\mu$ of the Poisson multiplication $\mu_0$ is a Lie formal deformation if the corresponding commutative associative multiplication is conserved, that is, if
$$\mu_0(x,y)+\mu_0(y,x)=\mu(x,y)+\mu(y,x)$$
for any $x,y$.
\end{definition}
As $\mu(x,y)=\mu_0(x,y)+\sum_{n\geq 1}t^n\mu_n(x,y)$, if $\mu$ is a Lie deformation of $\mu_0$, then
$$\mu(x,y)+\mu(y,x)=\mu_0(x,y)+\mu_0(y,x)+\sum_{n\geq 1}t^n(\mu_n(x,y)+\mu_n(y,x)).$$
So $$\sum_{n\geq 1}t^n(\mu_n(x,y)+\mu_n(y,x))=0$$
and
$$\mu_n(x,y)+\mu_n(y,x)=0$$
for any $n\geq 1.$ Each   bilinear maps $\mu_n$ is skew-symmetric. In particular $\mu_1$ is skew-symmetric
and $(\mu_1)_s=0$. As $\delta^2_P \mu_1 = 0$, Relation (\ref{deltaQ})
writes as
$$\delta^2_C \mu_1
+\mathcal{L}_H (\mu_1)+
\mathcal{L} _1(\mu_1)=0.$$
But, from  (\ref{chev}), $\delta^2_P \mu_1=0$ implies $\delta^2_C \mu_1=0.$ Thus we have $\mathcal{L}_H (\mu_1)+
\mathcal{L}_1(\mu_1)=0.$ 
Since $\mu_1$ is skew-symmetric:
$$\begin{array}{lll}
\mathcal{L}_H(\mu_1)(x,y,z)&=&\mu_1(x,y)\bullet z - x\bullet \mu_1(y,z) + \mu_1(x\bullet y,z)  -\mu_1(x,y\bullet z)  \\
&=& -\mu_1(x,y\bullet z) +  \mu_1(x,y) \bullet z+ y\bullet \mu_1(x,z) +\mu_1(x\bullet y,z) \\
&& - x\bullet \mu_1(y,z) -\mu_1(x,z) \bullet y \\
&=& \mathcal{L}_1(\mu_1)(x,y,z)+ \mathcal{L}_1(\mu_1)(y,z,x).
\end{array}
$$
So $$ \mathcal{L}_H (\mu_1)=\mathcal{L}_1 (\mu_1) \circ \Phi _{Id+c}.$$
We deduce that
$$\mathcal{L}_H (\mu_1)+ \mathcal{L}_1(\mu_1)=\mathcal{L}_1 (\mu_1) \circ \Phi _{2Id+c}$$ 
and 
$\mathcal{L}_H (\mu_1)+ \mathcal{L}_1(\mu_1)=0$ implies $\mathcal{L}_1 (\mu_1)=0.$

\begin{theorem}
If $\mu(x,y)=\mu_0(x,y)+\sum_{n\geq 1}t^n\mu_n(x,y)$ is a Lie deformation of the Poisson product $\mu_0$, then $\mu_1$ is a 
skew-symmetric map satisfying
$$
\left\{
\begin{array}{l}
\delta^2_C \mu_1=0,\\
\mathcal{L}_1(\mu_1)=0.
\end{array}
\right.
$$
\end{theorem}

\bigskip

Recall that  Poisson-Lichnerowicz cohomology \cite{lichne} is associated with the complex 
$$(\mathcal{C}^*_{PL}(\p,\p), \delta^*_{C})$$ where
the cochains are the skew-symmetric
multilinear maps $\p \times \cdots \times \p \rightarrow \p$ satisfying the Leibniz rule in each
of their arguments (such maps are called skew-symmetric multiderivations
of the algebra $\p$). The coboundary operators coincide with the Chevalley-Eilenberg coboundary operator denoted by 
$\delta_C^*$.
Of course $\mathcal{C}^n_{PL}(\p,\p)$ is a vector subspace of $\mathcal{C}^n_{P}(\p,\p)$. The previous theorem shows 
that if  $\varphi$ is a $2$-cochain of  $\mathcal{C}^2_{PL}(\p,\p)$, thus its classes of cohomology in 
$H^2_{PL}(\p,\p)$ and $H^2_{P}(\p,\p)$ are equal.

\bigskip

\noindent{\bf Remark.} Usually, only Lie deformations of Poisson algebras are considered. This is a consequence
of the classical problem of considering Poisson algebras on the associative commutative algebra
of differential functions on a manifold. In this context, the associative algebra is preserved
when we consider deformations of Poisson structures on this algebra, for example in problems of deformation quantization.
Moreover, such an associative structure is rigid, so it is not appropriate to consider deformations of this multiplication. As consequence, the corresponding deformation cohomology is the Poisson-Lichnerowicz cohomology. In our context, the deformation cohomology is given by a more general complex. In the next section, we will study the special non classical case where the Lie bracket is preserved, but we deform the associative product.

\subsection{Associative deformations of Poisson algebras}

\begin{definition}
We say that the formal deformation $\mu$ of the Poisson multiplication $\mu_0$ is an {\rm associative formal deformation}
if the corresponding Lie multiplication is conserved, that is, if
$$\mu_0(x,y)-\mu_0(y,x)=\mu(x,y)-\mu(y,x)$$
for any $x,y$.
\end{definition}
As $\mu(x,y)=\mu_0(x,y)+\sum_{n\geq 1}t^n\mu_n(x,y)$, if $\mu$ is an associative deformation of $\mu_0$, then
$$\mu(x,y)-\mu(y,x)=\mu_0(x,y)-\mu_0(y,x)+\sum_{n\geq 1}t^n(\mu_n(x,y)-\mu_n(y,x)).$$
Thus $$\sum_{n\geq 1}t^n(\mu_n(x,y)-\mu_n(y,x))=0$$
and
$$\mu_n(x,y)-\mu_n(y,x)=0$$
for any $n\geq 1.$ Each bilinear maps $\mu_n$ is symmetric. In particular $\mu_1$ is symmetric and $(\mu_1)_a=0$.
Since $\delta^2_P \mu_1 = 0$, Relation (\ref{deltaQ})
 writes as
$$
\mathcal{L}_C(\mu_1)+\delta^2_H \mu_1+
\mathcal{L}_2(\mu_1)=0.$$
But, from  (\ref{chev1}), $\delta^2_P \mu_1 = 0$ implies $\delta^2_H \mu_1=0.$ Thus
$$\mathcal{L}_C(\mu_1)+\mathcal{L}_2(\mu_1)=0.$$

\begin{lemma}
When $\varphi $ is a symmetric map with $\delta^2_H \varphi=0$,
$$
\begin{array}{lll}
\medskip
\mathcal{L}_C(\varphi )(x,y,z)&=& \left\{ \varphi (x,y),z \right\} + \left\{ \varphi (y,z),x \right\} +
 \left\{\varphi (z,x),y \right\}\\
\medskip
&& +\varphi ( \left\{ x,y \right\},z)+\varphi ( \left\{ y,z\right\} ,x)+\varphi (\left\{ z,x \right\} ,y), \\
\medskip
\mathcal{L}_2(\varphi )(x,y,z)&=& \left\{y, \varphi (x,z) \right\} - \left\{z, \varphi (x,y) \right\} +
 3 \varphi (x,\left\{z,y\right\}).
\end{array}$$
\end{lemma}
This is a direct consequence of the definition of $\mathcal{L}_C(\varphi _s)$ and $\mathcal{L}_2(\varphi _s)$
when $\varphi $ is a symmetric bilinear map, replacing $\mu_0(x,y)-\mu_0(y,x)$ by $2 \left\{ x,y\right\}.$

We deduce 
$$
\begin{array}{lll}
(\mathcal{L}_C (\mu_1)+
\mathcal{L}_2(\mu_1))(x,y,z)&=& 2\{\mu_1(x,y),z\}+\{\mu_1(y,z),x\}+\mu_1(\{x,y\},z)\\
&&+\mu_1(\{z,x\},y)+2\mu_1(\{z,y\},x)\\
&=& 2\{\mu_1(x,y),z\}-2\mu_1(\{y,z\},x)-2\mu_1(\{x,z\},y)\\
&&+\{\mu_1(y,z),x\}-\mu_1(\{y,x\},z)-\mu_1(\{z,x\},y)\\
&=& 2\Delta \mu_1 (x,y,z)+\Delta \mu_1 (y,z,x)
\end{array}
$$
with $$\Delta \mu_1 (x,y,z)=\{\mu_1(x,y),z\}-\mu_1(\{y,z\},x)-\mu_1(\{x,z\},y).$$
We deduce that  
$$(\mathcal{L}_C (\mu_1)+
\mathcal{L}_2(\mu_1)) =\Delta \mu_1 \circ \Phi _{2Id+c}.$$
But $\Phi _{2Id+c}$ is an invertible map on $\p^{\otimes^3}.$ Then 
$(\mathcal{L}_C (\mu_1)+
\mathcal{L}_2(\mu_1)) =0$ if and only if $$\Delta \mu_1(x,y,z) =\{ \mu_1 (x,y),z\} -\mu_1(\{ y,z\},x)-\mu_1(\{x,z \},y)=0.$$

\begin{definition}
Let $\p$ be a Poisson algebra and let $\{x,y\}$ be its Poisson bracket. A bilinear map $\varphi$ on $\p$ is called a Lie biderivation if
$$\{\varphi(x_1,x_2),x_3\}-\varphi(x_1,\{x_2,x_3\})-\varphi(\{x_1,x_3\},x_2)=0$$
for any $x_1,x_2,x_3 \in \p.$
\end{definition}
We deduce that $\mu_1$, which is a symmetric map, is a Lie biderivation.
\begin{theorem}
If $\mu(x,y)=\mu_0(x,y)+\sum_{n\geq 1}t^n\mu_n(x,y)$ is an associative deformation of the Poisson product $\mu_0$, then $\mu_1$ is a symmetric map such that
\begin{enumerate}
 \item $\delta^2_H \mu_1=0.$
\item $\mu_1$ is a Lie biderivation.
\end{enumerate}
\end{theorem}
In case of Lie deformation of the Poisson product $\mu_0$, we have seen that the relations concerning $\mu_1$ can be interpreted in terms
of Poisson-Lichnerowicz cohomology. We propose a similar approach for the Lie deformations of $\mu_0$.

\medskip

Recall that $x\bullet y$ the associative commutative product associated with the Poisson product $\mu_0$, that is $\displaystyle x\bullet y=\frac{\mu_0(x,y)+\mu_0(y,x)}{2}$,
\begin{lemma}
Let $\varphi$ be a symmetric bilinear map on $\p$ which is a Lie biderivation. If $\delta^2_H \varphi$ is the Hochschild coboundary operator, 
we have
$$
\delta^2_H \varphi(x_1,x_2,x_3)= x_1\bullet \varphi(x_2,x_3)-\varphi(x_1\bullet x_2,x_3)+ \varphi(x_1,x_2\bullet x_3)-\varphi(x_1,x_2)\bullet x_3,
$$
and 
$$
\{\delta^2_H \varphi(x_1,x_2,x_3),x_4\}=\delta^2_H \varphi(\{x_1,x_4\},x_2,x_3)+\delta^2_H \varphi(x_1,\{x_2,x_4\},x_3)
+\delta^2_H \varphi(x_1,x_2,\{x_3,x_4\})
$$
for any $x_1,x_2,x_3,x_4 \in \p.$
\end{lemma}

\noindent{\it Proof.} As $\varphi$ is a Lie biderivation, we have
$$\{\varphi(x_1,x_2),x_3\}-\varphi(x_1,\{x_2,x_3\})-\varphi(\{x_1,x_3\},x_2)=0.$$
Thus, using the definition of $\delta^2_H \varphi$, we obtain
$$
\begin{array}{ll}
\{\delta^2_H \varphi(x_1,x_2,x_3),x_4\} &= \{x_1\bullet \varphi(x_2,x_3),x_4\}-\{\varphi(x_1\bullet x_2,x_3),x_4\}+ \{\varphi(x_1,x_2\bullet x_3),x_4\}\\
&- \{\varphi(x_1,x_2)\bullet x_3,x_4\}\\
&= x_1\bullet \{\varphi(x_2,x_3),x_4\}-x_3\bullet \{\varphi(x_1,x_2),x_4\}+ \varphi(x_2,x_3)\bullet \{x_1,x_4\}\\
&- \varphi(x_1,x_2)\bullet \{x_3,x_4\}-\{\varphi(x_1\bullet x_2,x_3),x_4\}+ \{\varphi(x_1,x_2\bullet x_3),x_4\}
\end{array}
$$
As $\varphi$ is a Lie biderivation,
$$
\begin{array}{lll}
\{\delta^2_H \varphi(x_1,x_2,x_3),x_4\}&=& \! \!
x_1\bullet \varphi(\{x_2,x_4\},x_3)+x_1\bullet\varphi(x_2,\{x_3,x_4\})-x_3\bullet \varphi(\{x_1,x_4\},x_2)\\
&& \! \! -x_3\bullet\varphi(x_1,\{x_2,x_4\})+\varphi(x_2,x_3)\bullet \{x_1,x_4\} -\varphi(x_1,x_2)\bullet \{x_3,x_4\}\\
&&\! \! -\varphi(x_1 \bullet\{x_2,x_4\},x_3) -\varphi(x_2 \bullet\{x_1,x_4\},x_3) -\varphi(x_1 \bullet x_2,\{x_3,x_4\})\\
&&\! \! +\varphi(\{x_1,x_4\},x_2 \bullet x_3)+\varphi(x_1,x_2 \bullet\{x_3,x_4\})+\varphi(x_1,x_3 \bullet\{x_2,x_4\}). \\
\end{array}
$$
But
$$
\begin{array}{lll}
\delta^2_H \varphi(\{x_1,x_4\},x_2,x_3)&=& \! \!\{x_1,x_4\}\bullet \varphi(x_2,x_3) -\varphi( \{x_1,x_4\}\bullet x_2 ,x_3)+\varphi(\{x_1,x_4\},x_2 \bullet x_3)\\
&&- \varphi(\{x_1,x_4\},x_2)\bullet x_3.\\
\delta^2_H \varphi(x_1,\{x_2,x_4\},x_3)&=&\! \!x_1\bullet \varphi(\{x_2,x_4\},x_3) -\varphi( x_1\bullet \{x_2,x_4\} ,x_3)+\varphi(x_1,\{x_2,x_4\} \bullet x_3)\\
&&- \varphi(x_1,\{x_2,x_4\})\bullet x_3.\\
\delta^2_H \varphi(x_1,x_2,\{x_3,x_4\})&=&\! \!x_1\bullet \varphi(x_2,\{x_3,x_4\}) -\varphi( x_1\bullet x_2 ,\{x_3,x_4\})+\varphi(x_1,x_2 \bullet \{x_3,x_4\})\\
&&- \varphi(x_1,x_2)\bullet \{x_3,x_4\}.\\
\end{array}
$$
As the product $\bullet$ is commutative, we deduce
$$\{\delta^2_H \varphi(x_1,x_2,x_3),x_4\}=\delta^2_H \varphi(\{x_1,x_4\},x_2,x_3)+\delta^2_H \varphi(x_1,\{x_2,x_4\},x_3)+\delta^2_H \varphi(x_1,x_2,\{x_3,x_4\}).$$
Observe that the last identity is not a consequence of the symmetry of $\varphi$. It is satified for any bilinear Lie 
biderivation. Now, we can generalize these identities.
\medskip
\begin{definition}
Let $\phi$ be a $k$-linear map on $\p$. We say that $\phi$ is a Lie $k$-derivation if
$$\{\phi(x_1,\cdots,x_k),x_{k+1}\}=\sum_{i=1}^{k}\phi(x_1,\cdots,\{x_i,x_{k+1}\},\cdots,,x_k)$$
for any $x_1,\cdots,x_{k+1} \in \p$, where $\{x,y\}$ denotes the Lie bracket associated with the Poisson product.
\end{definition}
For example, from the previous lemma, if $\varphi$ is a Lie $2$-derivation (or biderivation), then $\delta^2_H \varphi$ is a Lie $3$-derivation.

\medskip

\noindent For any $(k-1)$-linear map on $\p$, let $\delta^{k-1}_H \varphi$ the $k$-linear map given by
$$
\begin{array}{lll}
\delta^{k-1}_H \varphi (x_1,\cdots,x_{k})& \! \! \!=& \! \! \! x_1\bullet\varphi (x_2,\cdots,x_{k})-\varphi (x_1\bullet x_2,\cdots,x_{k})+\varphi (x_1,x_2\bullet x_3,\cdots,x_{k})+\cdots\\
&& \! \! \! +(-1)^{k-1}\varphi (x_1,x_2,\cdots,x_{k-1} \bullet x_{k})+(-1)^{k}\varphi (x_1,x_2,\cdots,x_{k-1} )\bullet x_{k}.
\end{array}
$$
This operator is the coboundary operator of the Hochschild complex related to the associative operad
$\mathcal{A}ss$.
\begin{theorem}
If $\varphi$ is a Lie $k$-derivation of $\p$, then $\delta^k_H \varphi$ is a Lie $(k+1)$-derivation of $\p$.
\end{theorem}
\pf It is analogous to the proof detailed for $k=3$. It depends only of the symmetry of the associative product $x \bullet y$.

\medskip

\noindent Recall that a $k$-linear map $\varphi$ on a vector space is called commutative if it satisfies
$\varphi \circ \phi_{V_k}=0$ where 
$V_k=\sum_{\sigma \in  \Sigma _k}\varepsilon (\sigma)\sigma=0.$
\begin{lemma}
For any $k$-linear commutative map $\varphi$ on $\p$, the $(k+1)$-linear map $\delta^{k}_H \varphi$ is commutative.
\end{lemma}
\pf In fact, consider the first term of $\delta^{k}_H \varphi (x_1,\cdots,x_{k+1}),$ that is, 
$$x_1\bullet\varphi(x_2,\cdots,x_{k+1}).$$ 
We have $$\sum_{\sigma \in \Sigma^i_{k+1} }\varepsilon(\sigma) x_i \bullet \varphi(x_{\sigma(1)},\cdots ,x_{\sigma(i-1)},
x_{\sigma(i+1)},\cdots , x_{\sigma(k+1)})=0$$
because $\varphi$ is commutative, where  $\Sigma^i_{k+1}=\{ \sigma \in \Sigma_{k+1}, \sigma(i) =i\}.$ The same
trick vanishes the last terms, that is,
$$ \sum_{\sigma \in \Sigma{k+1}} \varphi (x_\sigma(1),x_\sigma(2),\cdots,x_{\sigma(k)} )\bullet x_{\sigma(k+1)}.$$
The terms in between  vanishes two by two when we compose with $\Phi_{V_k}.$

\medskip

\noindent Let $C^k_{PH}(\p,\p) $ be the vector space constituted by $k$-linear maps on $\p$ which are commutative and 
which are Lie $k$-derivations. From the previous result, the image of the $C^k_{PH}(\p,\p) $ by the map 
$\delta^k_H $ is contained in $C^{k+1}_{PH}(\p,\p) $. As these maps coincide with the coboundary operators of the 
 complex, we
 obtain a complex $(C^k_{PH}(\p,\p),\delta^k_H)$ whose associated cohomology is called the Poisson-Hochschild cohomology.
 \begin{theorem}
 Let $\p$ be a Poisson algebra whose (nonassociative) product is denoted $\mu_0$. For any associative deformation $\mu=\sum_{n\geq 0}t^i\mu_i$  of $\mu_0$, the linear term $\mu_1$ is a $2$-cocycle for the Poisson-Hochschild cohomology.
 \end{theorem}

 \medskip

 \subsection{Example: Poisson structures on rigid Lie algebras} Such  Poisson structures have been studied in 
\cite{Nico, Mic-Eli-Poisson}. We will study these structures in terms of Poisson-Hochschild cohomology. Consider, for example, the $3$-dimensional complex Poisson algebra given, in a basis $\{e_1,e_2,e_3\},$ by
 $$e_1e_2=2e_2, \, e_1e_3=-2e_3, \, e_2e_3=e_1.$$
 If $\{,\}$ and $\bullet$ denote respectively the Lie bracket and the commutative associative product attached with the Poisson product, we have
 $$\{e_1,e_2\}=2e_2, \, \{e_1,e_3\}=-2e_3, \, \{e_2,e_3\}=e_1$$
 and
 $$e_i\bullet e_j=0,$$
 for any $i,j.$
 If $\varphi$ is a Lie biderivation, it satisfies
 $$\{\varphi(e_i,e_j),e_k\}=\varphi(\{e_i,e_k\},e_j)+\varphi(\{e_j,e_k\},e_i).$$
 This implies $\varphi=0$ and the Poisson algebra is rigid.

 \section{Poisson cohomology} \label{7}

In this section, we describe relations between the coboundary operators $\delta^k _{P}$ of the Poisson cohomology 
(the operadic cohomology or the deformation cohomology)
 of a Poisson algebra $\p$ and the corresponding operators of the Poisson-Lichnerowicz and Poisson-Hochschild cohomology of $\p$.

 \subsection{The cases $k=0$ and $k=1$} 
\begin{itemize} 
\item $k=0.$ We put
$$\ H_P^{0}(\mathcal{P},\mathcal{P})= \left\{ X \in \mathcal{P} \, {\mbox{\rm such that}} \,
\forall Y \in \mathcal{P}, X\cdot Y=0 \right\} .$$

\item $k=1.$ For $f \in End (\mathcal{P},\mathcal{P}),$ we put
$$\delta^1_{P}f(X,Y)=f(X)\cdot Y+X \cdot f(Y)-f(X\cdot Y) $$
for any $ X,Y \in \mathcal{P}.$ Then we have

$$ H_P^{1}(\mathcal{P},\mathcal{P})= \,  \! H_C^{1}(\frak{g}_{\mathcal{P}},\frak{g}_{\mathcal{P}})
\cap \,
 \! H_H^{1}( \mathcal{A}_{\mathcal{P}},\mathcal{A}_{\mathcal{P}}).$$
\end{itemize}

\subsection{Description of  $\delta^2 _{P}$}

In Section 4, we have seen that
$$
\begin{array}{ll}
\delta^2_{P}\varphi (x,y,z)= & 3\varphi( x \cdot y,z)-3\varphi (x,y\cdot z)-\varphi( x \cdot z,y)-
\varphi( y \cdot z,x)
\\ & +\varphi( y \cdot x,z)+\varphi( z \cdot x,y) +3\varphi(x,y)\cdot z-3 x \cdot \varphi (y,z)
\\ & -\varphi (x,z) \cdot y- \varphi (y,z)\cdot x + \varphi (y,x) \cdot z
+\varphi (z,x) \cdot y
\end{array}
$$
and
$$
\begin{array}{l}
\delta^2_P \varphi =2(\delta^2_C \varphi_a+
\mathcal{L}_C (\varphi_s)+\delta^2_H \varphi_s+\mathcal{L}_H (\varphi_a)+
\mathcal{L}_1(\varphi_a)+\mathcal{L}_2({\varphi_s})).
\end{array}
$$

Let us compare this operator with the corresponding Poisson-Lichnerowicz and Poisson-Hochschild ones. 

\noindent{\bf Example.} Assume that the Poisson product is skew-symmetric. Then $\{x,y\}=x \cdot y$
and $x \bullet y=0.$
If $\varphi  \in \mathcal{C}_P^2(\mathcal{P},\mathcal{P})$ is also skew-symmetric, then
$$\begin{array}{rl}
\delta_{P}^2 \varphi(x,y,z)= & 2\varphi(x \cdot y,z)+2\varphi (y\cdot z,x)-2\varphi (x\cdot z,y)\\
& +2\varphi (x,y)\cdot z+2\varphi (y,z) \cdot x-2\varphi (x,z)\cdot y \\
= &\delta_{PL}^2 \varphi(x,y,z),
\end{array}$$
that is, the coboundary operator of the Poisson-Lichnerowicz cohomology. 
\medskip

Let $\varphi_s$ and $\varphi_a$ be the symmetric and skew-symmetric parts of
$\varphi  \in \mathcal{C}_P^2(\mathcal{P},\mathcal{P}).$ The results of the previous sections imply:
\begin{theorem}
\label{theo}
Let $\varphi $ be in $ C_P^2(\mathcal{P},\mathcal{P}),$   $\varphi_s$ and $\varphi_a$ be its
symmetric and skew-symmetric parts. Then the following propositions are equivalent:

1. $\delta ^2_{P}\varphi =0.$

\medskip

2. $\left\{
\begin{array}{l}
i) \ \delta ^2_C\varphi _a= 0, \ \delta ^2_H\varphi _s = 0, \\
ii) \ \mathcal{L}_C (\varphi_s)+\mathcal{L}_H (\varphi_a)
 + \mathcal{L}_1(\varphi_a)+ \mathcal{L}_2(\varphi_s) =0.
\end{array}
\right. $
\end{theorem}
\medskip

\noindent{\bf Applications.}

\noindent Suppose that $\varphi $ is skew-symmetric. Then $\varphi =\varphi _a$ and $\varphi _s=0$.
Then $\delta ^2_{P}\varphi =0$ if and only if $\delta ^2_C\varphi = 0$
and $  \mathcal{L}_H (\varphi)
 + \mathcal{L}_1(\varphi) =0.$ Morever if we suppose than $\varphi $ is a biderivation
on each argument,
 that is, $\mathcal{L}_1(\varphi) =0$, then $\delta ^2_{P}\varphi =0$
 if and only if $\mathcal{L}_H (\varphi) =0.$ But we have seen in Section 3 that
 $$\mathcal{L}_H (\varphi) =\mathcal{L}_1 (\varphi) \circ \Phi_{Id+c}.$$
  Thus $\mathcal{L}_H ( \varphi) =0$ as soon as $\mathcal{L}_1(\varphi) =0.$
\begin{proposition}
Let $\varphi $ be a skew-symmetric map which is a biderivation, that is $\varphi $ is a Poisson-Lichnerowicz
$2$-cochain. Then $\varphi \in Z^2_{PL}(\mathcal{P},\mathcal{P})$ if and only if
$\varphi \in Z^2_{P}(\mathcal{P},\mathcal{P})$.
\end{proposition}
Similarly, if $\varphi $ is symmetric, then $\delta ^2_{P}\varphi =0$
if and only if $\ \delta ^2_H\varphi = 0$
and $\mathcal{L}_C ( \varphi)
 + \mathcal{L}_2(\varphi) =0.$ 
If $\varphi $ be a skew-symmetric map which is a Lie biderivation, that is, if $\varphi $ is a Poisson-Hochschild
$2$-cochain, then $\varphi \in Z^2_{PH}(\mathcal{P},\mathcal{P})$ if and only if
$\varphi \in Z^2_{P}(\mathcal{P},\mathcal{P})$.

\subsection{The case $k \geq 3$}

Let $\p$ be a Poisson algebra and $H_{def}^*(\p)$  or  $H_{P}^*(\p,\p)$ its operadic cohomology. 
We propose here to describe  $H_{P}^n(\p,\p)$ for $n\geq 3.$
Let $\varphi$ be a $n$-cochain of  $C_{P}^n(\p,\p),$ that is, 
a $n$-linear map on $\p.$ Its skew-symmetric part is the skew-symmetric $n$-linear map 
$$\varphi_a=\frac{1}{n!} \varphi \circ \Phi_{V_n}$$
with $V_n=\sum_{\sigma \in \Sigma_n } \varepsilon(\sigma)\sigma;$ its symmetric part is the symmetric $n$-linear map
$$\varphi_s=\frac{1}{n!} \varphi \circ \Phi_{W_n}$$
with $W_n=\sum_{\sigma \in \Sigma_n } \sigma.$
We denote by $\delta_P^n,\delta_C^n$ and $\delta_H^n$ respectively the coboundary 
operators associated with the Poisson cohomology of $\p$, the Chevalley-Eilenberg cohomology 
of $\frak{g}_{\p}$ and the Hochschild cohomology of $\mathcal{A}_{\p}.$

The formulae (\ref{chev}) and (\ref{chev1}) can be generalized as follows
\begin{eqnarray}
\label{chev(n)}
2(n+1)!\delta_C^n\varphi_a=\delta_{p}^n \varphi \circ \Phi_{V_n},
\end{eqnarray}
\begin{eqnarray}
\label{chev1(n)}
2(n+1)!\delta_H^n\varphi_s=\delta_{p}^n \varphi \circ \Phi_{U_{H,n}},
\end{eqnarray}
where $U_{H,n}=\sum_{\sigma \in \Sigma_{1,n}}\sigma +(-1)^n \sum_{\sigma \in \Sigma_{n,n}}\sigma$ 
with $ \Sigma_{i,n}=\left\{  \sigma \in \Sigma_n , \sigma(1)=i  \right\}.$

\begin{proposition}
Let $\varphi$ be a $n$-cochain of the Poisson complex of the Poisson algebra $\p$. Then
$$\delta_P^n\varphi=0 \Rightarrow \left\{
\begin{array}{l}
\delta_C^n \varphi_a =0, \\
\delta_H^n \varphi_s =0.
\end{array}
\right.$$
\end{proposition}
Let us consider $\mathcal{L}_{1,n}$ acting on the skew-symmetric $n$-linear map by
$$
\begin{array}{lll}
\medskip
2(n-1)!\mathcal{L}_{1,n}\varphi_a&=& \sum_{\sigma^{-1} \in \Sigma_{i,i+1,n}} \varepsilon(\sigma) \varphi \circ_{\sigma^{-1}(1)} \mu
 \circ \Phi_{(Id+\tau_{12}) \circ \sigma}\\
\medskip
&& +(-1)^{n-1}  \sum_{\sigma^{-1} \in \Sigma_{n,n}} \varepsilon(\sigma) \mu \circ_{1} \varphi \circ \Phi_{(Id+\tau_{12})\circ\sigma}\\
&& -  \sum_{\sigma^{-1} \in \Sigma_{1,n}} \varepsilon(\sigma) \mu \circ_{2} \varphi \circ \Phi_{(Id+\tau_{12})\circ\sigma}\\
\end{array}$$
where $  \Sigma_{i,i+1,n}=\left\{ \sigma \in \Sigma_n, \sigma(1)=i,\sigma(2)=i+1\right\}.$

\begin{lemma}
$\varphi_a$ is a skew-symmetric $n$-derivation, that is, a skew-symmetric $n$-linear map which is a derivation 
for the associative product $x \bullet y$ on each argument, if and only if $\mathcal{L}_{1,n}\varphi_a=0.$
\end{lemma}
Now we define the operator  $\mathcal{L}_{H,n}$which acts on the  the skew-symmetric $n$-linear map by 
$$ \mathcal{L}_{H,n}\varphi_a=\mathcal{L}_{1,n}\varphi_a \circ \Phi_{Id+c_n+c_n^2+ \cdots +c_n^{n-2}}$$
where $c_n \in \Sigma_n$ is the cycle $(1,2,\cdots,n).$

\begin{proposition}
Let $\varphi$ be a skew-symmetric linear map on $\p^{\otimes^n}.$ Then 
$\delta_P^n\varphi=0$ if and only if $\delta_C^n \varphi=0 $ and  $ \mathcal{L}_{1,n}\varphi=0. $
\end{proposition}

We find again the classical result: the associative deformations of a Poisson algebra are parametrized by the Poisson-Lichnerowciz cohomology.

\medskip

Assume now that $\varphi$ is a symmetric $n$-linear map. We have seen that:
$$\delta_P^n\varphi=0 \Rightarrow \delta_H^n\varphi_s=\delta_H^n\varphi=0.$$
Consider the operator $\nabla^n$ acting on the symmetric $n$-linear maps by:
$$\begin{array}{lll}
\nabla^n \varphi_s (x_1, \cdots, x_{n+1})&=&\left\lbrace \varphi(x_1, \cdots , x_n),x_{n+1} \right\rbrace - 
\varphi (\left\lbrace x_1,x_{n+1} \right\rbrace ,x_2  \cdots x_n)\\
& & -\varphi (x_1,\left\lbrace x_2,x_{n+1} \right\rbrace ,x_3  \cdots x_n)- \cdots \\
& & -\varphi (x_1,x_2,\cdots, x_{n-1},\left\lbrace x_n,x_{n+1}\right\rbrace).\\
\end{array}$$
Then $\varphi=\varphi_s$ is a Lie $n$-derivation if and only if $\nabla^n\varphi_s=0.$

Now we consider the following operator acting also on the symmetric $n$-linear maps by:
$$\begin{array}{lll}
\mathcal{L}_C^n\varphi_s &=& \mu \circ_1\varphi \circ \Phi_{-c+c^2+\cdots+(-1)^{n+1}c^{n+1}} 
 +\mu \circ_2\varphi \circ \Phi_{Id-c+c^2+\cdots+(-1)^{n}c^{n}} \\
& & +\varphi \circ_1 \mu \circ \Phi_{\sum_{1 \leq i,j \leq n+1}(-1)^{i+j+1}c_{ij}}
\end{array}$$
where $c_{ij}$ is the permutation 
$\left(
\begin{array}{llllclcll}
1 & 2 & 3 & \cdots & \cdots    & \cdots & \cdots & \cdots & n+1 \\
i & j & 1 & \cdots & \check{i} & \cdots & \check{j} & \cdots & n+1
\end{array}.
\right)$
and $\mathcal{L}_2^n\varphi_s $ defined by:
$$\mathcal{L}_C^n\varphi_s +\mathcal{L}_2^n\varphi_s =\nabla^n \varphi_s \circ \Phi_u$$
with $u \in \mathbb{K}[\Sigma_{n}]$ equal to $\tau_{12}+\tau{13}+\cdots +\tau_{1n}.$
Since $\Phi_u$ is invertible, the equation $\mathcal{L}_C^n\varphi_s +\mathcal{L}_2^n\varphi_s =0$ implies
$\nabla^n \varphi_s = 0$ and we find that the Poisson-Hochschild cohomology coincides with the Poisson cohomology when
$\varphi=\varphi_s.$

\end{document}